\documentclass[10pt]{article}

\usepackage{epsfig}

\date{}

\title{
{
\Large \bf Hybrid Neural Networks for Frequency Estimation of Unevenly Sampled Data}}
\author{\large Roberto Tagliaferri$^{1}$, Angelo Ciaramella$^{1}$, Leopoldo Milano$^{2}$\\ \large and Fabrizio Barone$^{2}$ \\
$^{1}${\large DMI, Universit\`{a} di Salerno, and INFM unit\`{a} di Salerno,}
\\
{\large via S. Allende, 84081 Baronissi (SA) Italia }\\
$^{2}${\large Dip. di Scienze Fisiche, Universit\`{a} di Napoli ''Federico
II''}\\
{\large and INFN, sez. Napoli, via Cintia, I-80126 Napoli Italia}}

\begin{document}

\vspace{-2truecm}
\twocolumn
\columnsep 0.8truecm 
\maketitle	 

\thispagestyle{empty}


\begin{center}
{\bf \large Abstract}
\end{center}

{\it
\noindent 
In this paper we present a hybrid system composed by a neural network based
estimator system and genetic algorithms. It uses an unsupervised Hebbian
nonlinear neural algorithm to extract the principal components which, in
turn, are used by the MUSIC frequency estimator algorithm to extract the
frequencies.We generalize this me\-thod to avoid an interpolation
preprocessing step and to improve the performance by using a new stop
criterion to avoid overfitting. Furthermore genetic algorithms are used to
optimize the neural net weight initialization.
\\
\\
\noindent 
The experimental results are obtained comparing our me\-tho\-do\-logy with the
others known in literature on a Cepheid star light curve.
}
\\


\begin{center}
{\bf \large 1. Introduction }
\end{center}

\noindent 
Periodicity analysis of unevenly collected data is a relevant issue in
several scientific fields. Classical spectral analysis methods are
unsatisfactory to solve the problem. In this paper we present a neural
network based estimator system which performs well the frequency extraction
in unevenly sampled signals. It uses an unsupervised Hebbian nonlinear
neural algorithm to extract the principal components of the signal
auto-correlation matrix, which, in turn, are used by the MUSIC frequency
estimator algorithm to extract the frequencies \cite{Khar94}, \cite
{Tagliaferri 99}. We generalize this method to avoid an interpolation
preprocessing step, which generally adds high noise to the signal, and
improve the system performance by using a new stop criterion to avoid
overfitting problems. Furthermore, genetic algorithms are used to optimize
the neural net weight initialization, so improving its convergence. The
experimental results are obtained comparing our methodology with the others
known in literature \cite{Kay},\cite{Opp},\cite{Lomb76},\cite{Scargle},\cite
{Ferraz-Mello},\cite{Proakis}. 
\\

\begin{center}
{\bf \large 2. Evenly and unevenly sampled data}
\end{center}

\noindent
In what follows, we assume $x$ to be a physical variable measured at
discrete times $t_{i}$. ${x(t_{i})}$ can be written as the sum of the signal 
$x_{s}$ and random errors $R$: $x_{i}=x(t_{i})=x_{s}(t_{i})+R(t_{i})$. The
problem we are dealing with is how to estimate fundamental frequencies which
may be present in the signal $x_{s}(t_{i})$ \cite{Deeming}, \cite{Kay}.
\\
\\
\noindent 
If $X$ is measured at uniform time steps (even sampling) \cite{Horne},\cite
{Scargle} there are a lot of tools to effectively solve the problem which
are based on Fourier analysis \cite{Kay},\cite{Opp}. These methods,
however, are usually unreliable for unevenly sampled data. For instance, the
typical approach of resampling the data into an evenly sampled sequence,
through interpolation, introduces a strong amplification of the noise which
affects the effectiveness of all Fourier based techniques which are strongly
dependent on the noise level \cite{Horowitz}.
\\
\\
\noindent 
To solve the problem of unevenly sampled data, we consider two classes of
spectral estimators:

Spectral estimators based on Fourier Transform (Least Squares methods);

Spectral estimators based on the eigen\-va\-lues and eigen\-vec\-tors of the
covariance matrix (Maximum Likelihood methods).
\\
\\
\noindent 
Classic Perio\-do\-gram \cite{Kay},\cite{Opp}, Lomb's Perio\-do\-gram \cite{Lomb76}, Scargle's Periodogram \cite{Scargle}, DCD\-FT \cite{Ferraz-Mello} are the methods of the first class that we use, while MUSIC 
\cite{Kay},\cite{Opp} and ES\-PRIT \cite{Roy},\cite{Proakis} belong to
the second class.
\\
\\
\noindent 
The methods based on the covariance matrix are more recent and have great
potentiality. Starting by this consideration, we develop a method based on
the MU\-SIC estimator. It is compared with classical methods to evidence the
perfomance.
\\

\begin{center}
{\bf \large 3. The Neural Estimator}
\end{center}

\noindent 
In the last years several papers dealed with learning in PCA neural nets 
\cite{Oja82},\cite{Oja91},\cite{sanger},\cite{Khar94},\cite{Chen98}, \cite
{Tagliaferri 99} finding advantages, problems and difficulties of such
neural networks. In what follows, we shall use a robust hierarchical
learning algorithm because it has been experimentally shown that it is the
best performing in our problem \cite{Tagliaferri 99}.
\\
\\
\noindent 
Our neural estimator (ne) can be summarized as follows:

\begin{itemize}
\item[1]  Preprocessing: calculate and subtract the average pattern to
obtain zero mean process with unity variance.
\end{itemize}

Interpolate input data if it is the case.

\begin{itemize}
\item[2]  Train the neural network.

\item[3]  Calculate the frequencies estimation by using the frequency
estimator MUSIC. 
\end{itemize}

\noindent 
MUSIC takes as input the weight matrix co\-lumns of the neural network after
the learning. The estimated signal frequencies are obtained as the peak
locations of the function of the following equation \cite{Kay}: 
\[
P_{MUSIC}=\frac{1}{M-\sum_{i=1}^{M}|\mathbf{e}_{f}^{H}\mathbf{w}(i)|^{2}}
\]

\noindent 
where $\mathbf{w}(i)$ is the $i-$th neural network weight vector after
learning, and $\mathbf{e}_{f}^{H}$ is the vector 
\[\mathbf{e}_{f}^{H}=[1,e_{f}^{j2\pi ft_{0}},\ldots ,e_{f}^{j2\pi ft_{(L-1)}}]^{H}
\]
\noindent
where $\left\{ t_{0},t_{1},...,t_{\left( L-1\right) }\right\} $ are the
first $L$ components of the temporal coordinates of the uneven signal \cite
{Tagliaferri 99}.
\\
\\
\noindent 
When $f$ is the frequency of the $i-$th sinusoidal component, $f=f_{i}$, we
have $\mathbf{e}=\mathbf{e}_{i}$ and $P_{MUSIC}\rightarrow \infty $. In
practice we have a peak near and in corrispondence of the component
frequency. Estimates are related to the highest peaks
\\
\\
\noindent 
Furthermore, to optimize the performance of the PCA neural networks, we stop
the learning process when 

\[
\sum_{i=1}^{p}|\mathbf{e}_{f}^{H}\mathbf{w}(i)|^{2}<M\qquad \forall f,
\] 
so avoiding overfitting problems. In fact,
leaving the stop condition used in the ne causes to the ne to find
periodicities not present in the signal, while the new condition preserves
it from this problem.
\\
\\
\noindent 
Since sometimes the weight inizialization can lead to local maxima of the
objective function, here we propose to use genetic algorithms for this aim.
\\
\\
\noindent
We use a real encoding of the net parameters \cite{Lin} in a string
(chromosome), and then we run the genetic algorithms with a fitness
function, which is the system objective function. 
\\

\begin{center}
{\bf \large 4. Experimental Results }
\end{center}

\noindent 
Many experiments on synthetic and real signals were made \cite{Tagliaferri
99}, and in this paper we presents the results obtained with one specifc
real signal, which highlights the main features of our problem.
\\
\\
\noindent 
The signal is related to the Cepheid T Mon \cite{Moffet}. The sequence was
obtained with the photometric technique BVRI and the sampling made from April
1977 to December 1979. The light curve is composed by $24$ samples, and a
period of $27.02^{d}$, as shown in figure 1. In this case, the parameters
of the ne are: $N=10$, $p=2$, $\alpha =5$, $\mu =0.001$. The estimated
frequency interval is $\left[ 0(1/JD),0.5(1/JD)\right] $. The estimated
frequency is $0.020$ (1/JD) (see figure 2). By using the genetic algorithms (gne)we improve the system performance reaching the true frequency of $0.034$ (1/JD) (see figure 3). Lomb's Periodogram finds the correct frequency but it generates several spurious peaks (see figure 4). For what concerns DCDFT, the frequency interval must be changed because it is too much sensitive to very low frequencies (the new interval becomes 
$[ 0.01(1/JD), $\\$ 0.5(1/JD)]$) 
(see figure 5). 
Finally ESPRIT does not work at all finding a frequency of $0.23$ (1/JD) (see figure 6).
\\
\\
\noindent 
In conclusion, only the Lomb's Periodogram and our gne are in agreement with the right periodicity, but the former showing several spurious pe\-aks.
\\

\begin{figure}[tbp]
\vbox{
\hbox{
\psfig{figure=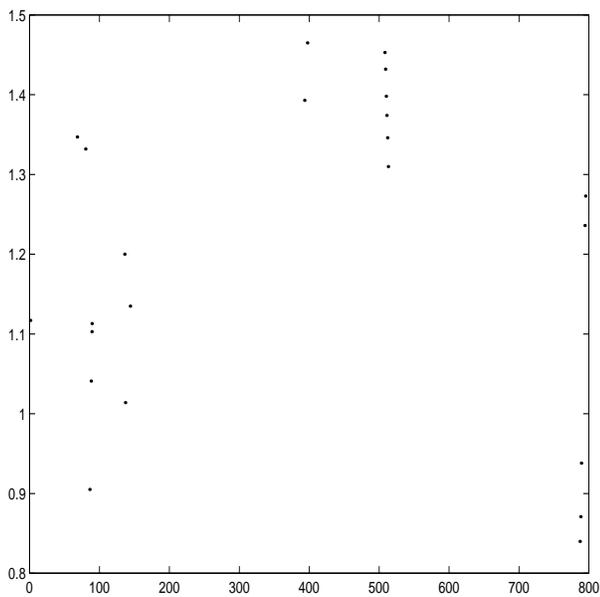,width=8cm,height=8cm}
}}
\caption{Light curve of T Mon
}
\end{figure}

\begin{figure}[tbp]
\vbox{
\hbox{
\psfig{figure=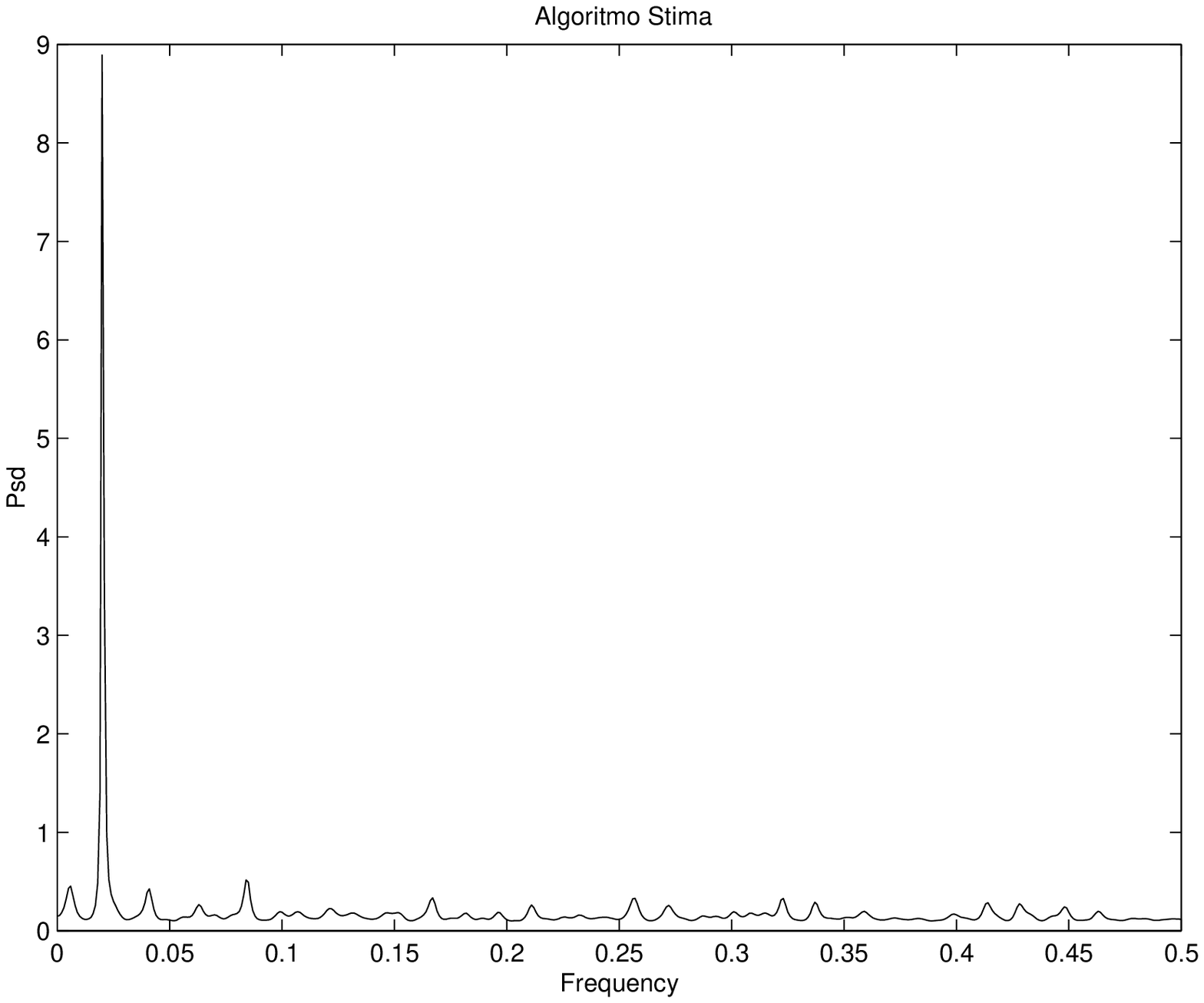,width=8cm,height=8cm}
}}
\caption{ne estimate of the fundamental frequency of T Mon
}
\end{figure}

\begin{figure}[tbp]
\vbox{
\hbox{
\psfig{figure=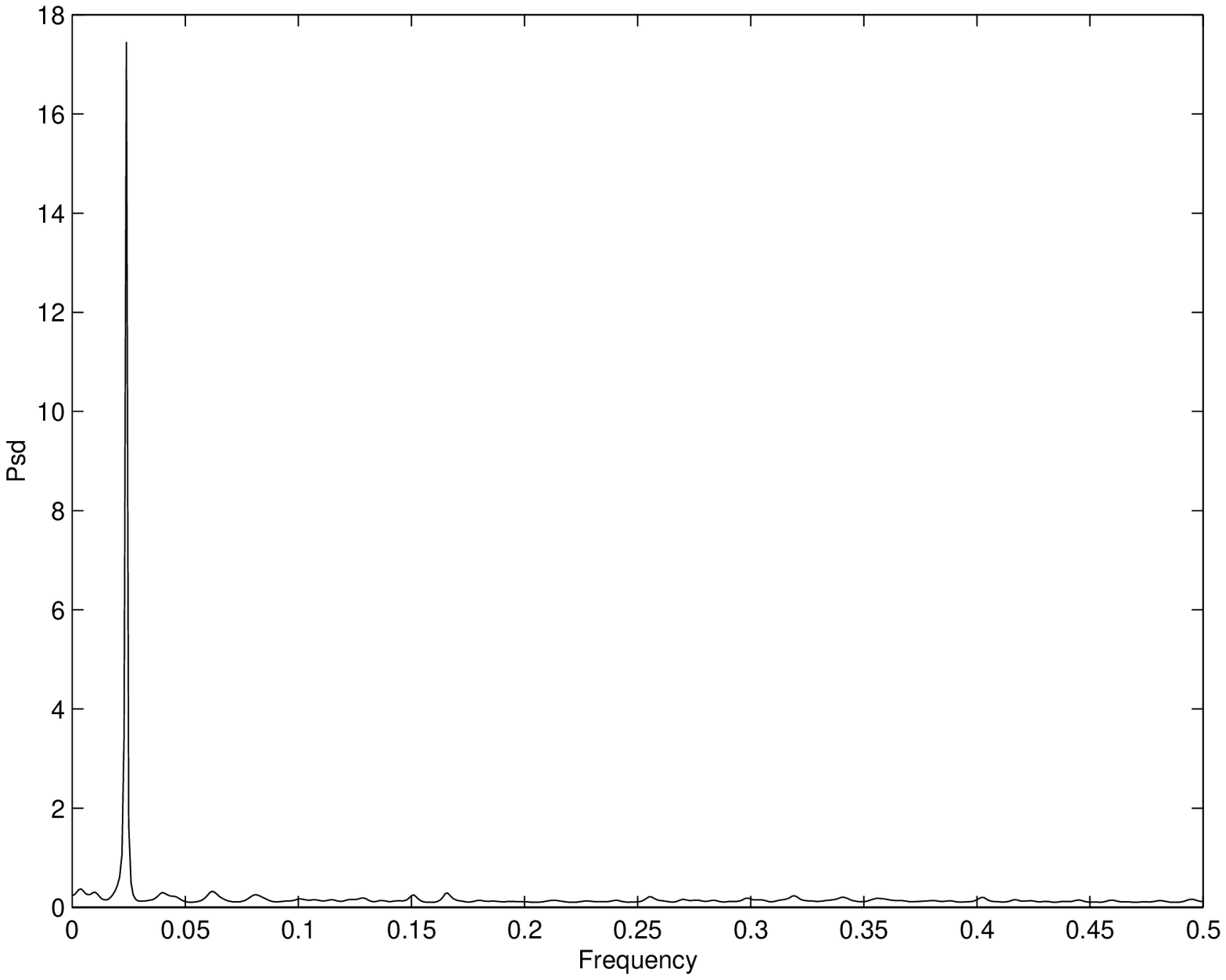,width=8cm,height=8cm}
}}
\caption{gne estimate of the fundamental frequency of T Mon
}
\end{figure}

\begin{figure}[tbp]
\vbox{
\hbox{
\psfig{figure=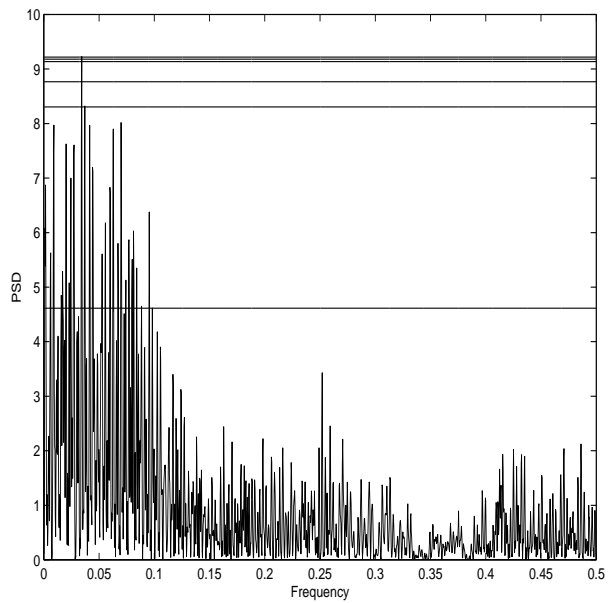,width=8cm,height=8cm}
}}
\caption{Lomb's Periodogram estimate of the fundamental frequency of T Mon
}
\end{figure}

\begin{figure}[tbp]
\vbox{
\hbox{
\psfig{figure=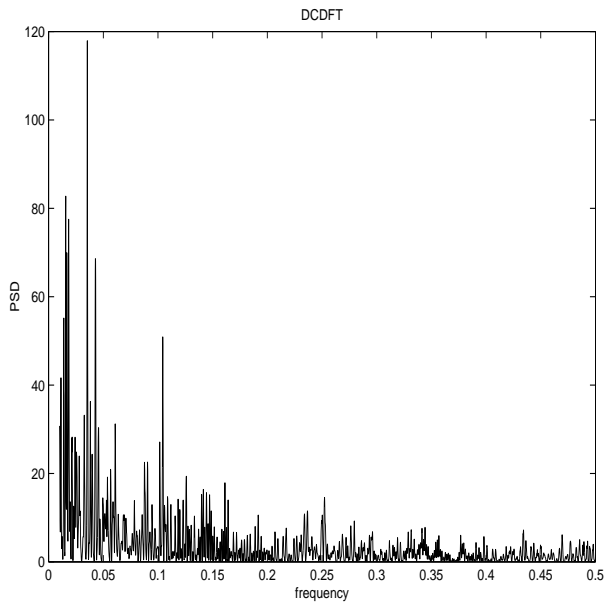,width=8cm,height=8cm}
}}
\caption{DCDFT estimate of the fundamental frequency of T Mon
}
\end{figure}

\begin{figure}[tbp]
\vbox{
\hbox{
\psfig{figure=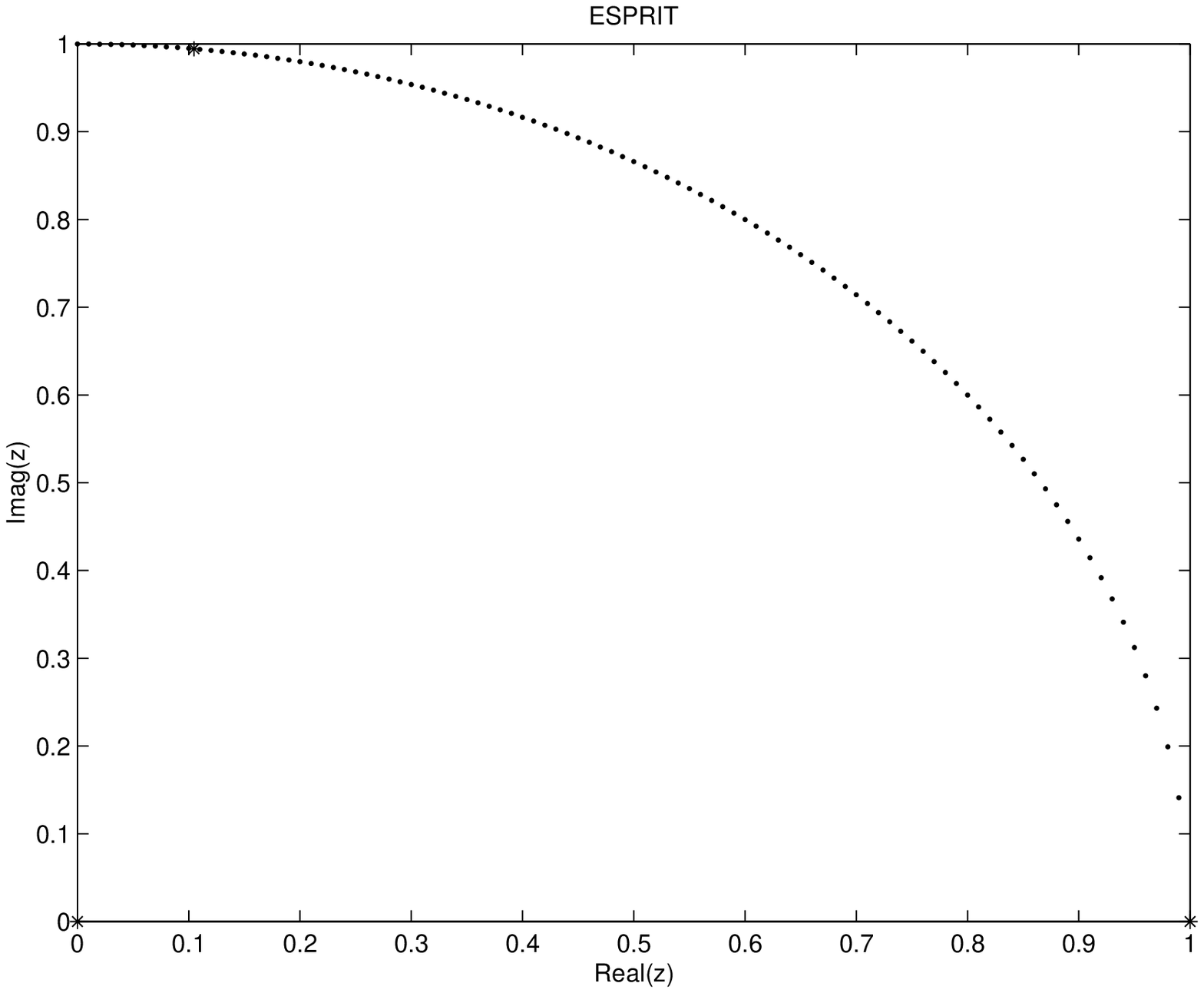,width=8cm,height=8cm}
}}
\caption{ESPRIT estimate of the fundamental frequency of T Mon
}
\end{figure}

\begin{center}
{\bf \large 5. Concluding Remarks}
\end{center}

\noindent 
In this paper we have illustrated an improved technique based on PCA neural
networks and MUSIC\ to estimate the frequency of unevenly sampled data. It
has been shown that it obtains good results on real data (here we used the
Cepheid T Mon light curve) compared with other well known methods. A further
improvement can be obtain by using filters to extract and identify one
frequency at each time when dealing with multi-frequency signals. In \cite
{Tagliaferribis}, we used our system to detect the Milankovic' frequencies
from a stratigrafic record. In that case the best performance was found when
we extracted one frequency at each time and eliminating it with a FIR
filtering.
\\

\begin{center}
{\bf \large Acknowledgements }
\end{center}

\noindent
The paper has been partially supported by IIASS ``E.\ R. Caianiello" and by MURST 40\%.
\\

\end{document}